\documentclass[scrartcl,aps,prl,reprint,superscriptaddress,amsmath,showpacs,amsfonts,amssymb,twocolumn,bibliography,longbibliography]{revtex4-1}

\usepackage{graphicx}
\usepackage{bm}
\usepackage{color}
\usepackage{MnSymbol}
\usepackage{enumerate}
\usepackage{bbold}
\usepackage{subfigure}
\usepackage{units}
\usepackage{gensymb}
\usepackage{setspace}
\usepackage{lineno}
\usepackage{layouts}

\def\be{\begin{equation}}
\def\ee{\end{equation}}

\begin{document}

\title{Reconciling and Validating the Ashworth-Davies Doppler Shifts of a Translating Mirror}

\author{Ziv Roi-Cohen}
\affiliation{Racah Institute of Physics, The Hebrew University of Jerusalem, Jerusalem, Israel, 91904}
\author{Merav Kahn}
\affiliation{Racah Institute of Physics, The Hebrew University of Jerusalem, Jerusalem, Israel, 91904}
\author{Nadav Katz}
\affiliation{Racah Institute of Physics, The Hebrew University of Jerusalem, Jerusalem, Israel, 91904}
\author{Stefania Residori}
\affiliation{HOASYS SAS, 120 route des Macarons, 06560 Valbonne, France} 
\author{Umberto Bortolozzo}
\affiliation{HOASYS SAS, 120 route des Macarons, 06560 Valbonne, France}
\author{John C. Howell}
\affiliation{Racah Institute of Physics, The Hebrew University of Jerusalem, Jerusalem, Israel, 91904}
\affiliation{Institute for Quantum Studies, Chapman University, 1 University Drive, Orange, CA 92866, USA}

\begin{abstract}
We simplify the Ashworth-Davies special relativistic theory of a uniformly translating mirror with an arbitrary angle of incidence and direction of propagation in the non-relativistic limit.  We show that it is in good agreement with a more intuitive derivation that only considers the constancy of the speed of light.  We experimentally confirm the theory predictions with phase-insensitive frequency measurements using a liquid crystal light valve.
\end{abstract}

\maketitle

\date{\today}
\section{Introduction}
A Doppler shift is the frequency differential of a wave when a source and detector are in relative motion.  Harnessing the Doppler effect has brought about great gains in scientific, engineering and society at large.  Doppler shifts are used extensively in astronomy \cite{butler1996attaining,sarre2006diffuse,deming2009discovery}, remote weather monitoring \cite{atlas1973doppler,bringi2001polarimetric,doviak2006doppler}, non-invasive medical diagnostics \cite{li2013review, lee2014monitoring, taylor1987clinical} and laser velocimetry (fluid flow) \cite{pedersen2003flow, dolan2010accuracy} to name a few applications.

Einstein's derivation of the Doppler shift of light from a uniformly translating mirror considered only the Doppler shift resulting from light reflected at an oblique angle \cite{einstein1905elektrodynamik}. While the angle of incidence of the beam relative to the surface normal was arbitrary, the direction of mirror propagation was in the same direction as the surface normal.  

A derivation of the Doppler shift that considers both arbitrary mirror propagation direction and arbitrary incidence angle was derived by Ashworth and Davies \cite{ashworth1976doppler}.  Follow-on experiments demonstrated the intended prediction from the theory namely that there is no Doppler effect for transversely moving mirrors \cite{jennison1974reflection,davies1977reflection}. However, no one has experimentally verified the results for the general case, even for nonrelativistic mirror velocities.

Understanding the general case of a Doppler shift from a moving mirror is important for our previous work on a Doppler-based gyroscope (see \cite{GyroPRL}). In that work, it was shown that Doppler shifts are fundamental in passive gyroscopes.  However, it is impossible to derive a generalized theory about the Doppler gyroscope without an understanding of the Doppler shifts result from the movement of the mirror with respect to its surface normal.  

When considering a rotating interferometer, each mirror in the interferometer can move in a nontrivial direction, depending on its position relative to the axis of rotation. To calculate the difference in frequency shift between the two optical paths in the interferometer, we need to understand what is the frequency shift that results from the reflection of each individual mirror.

Here, we show that the Ashworth-Davies result in the nonrelativistic domain can be written in an intuitive form that reconciles with a simple modification of a Doppler shift model by Ghurchinovski \cite{gjurchinovski2005doppler}. We verify the predictions of this derivation experimentally using a hypersensitive differential frequency measurement in a liquid crystal light valve \cite{bortolozzo2013precision}.  

\begin{figure}[htp]
\includegraphics[width=.45\textwidth]{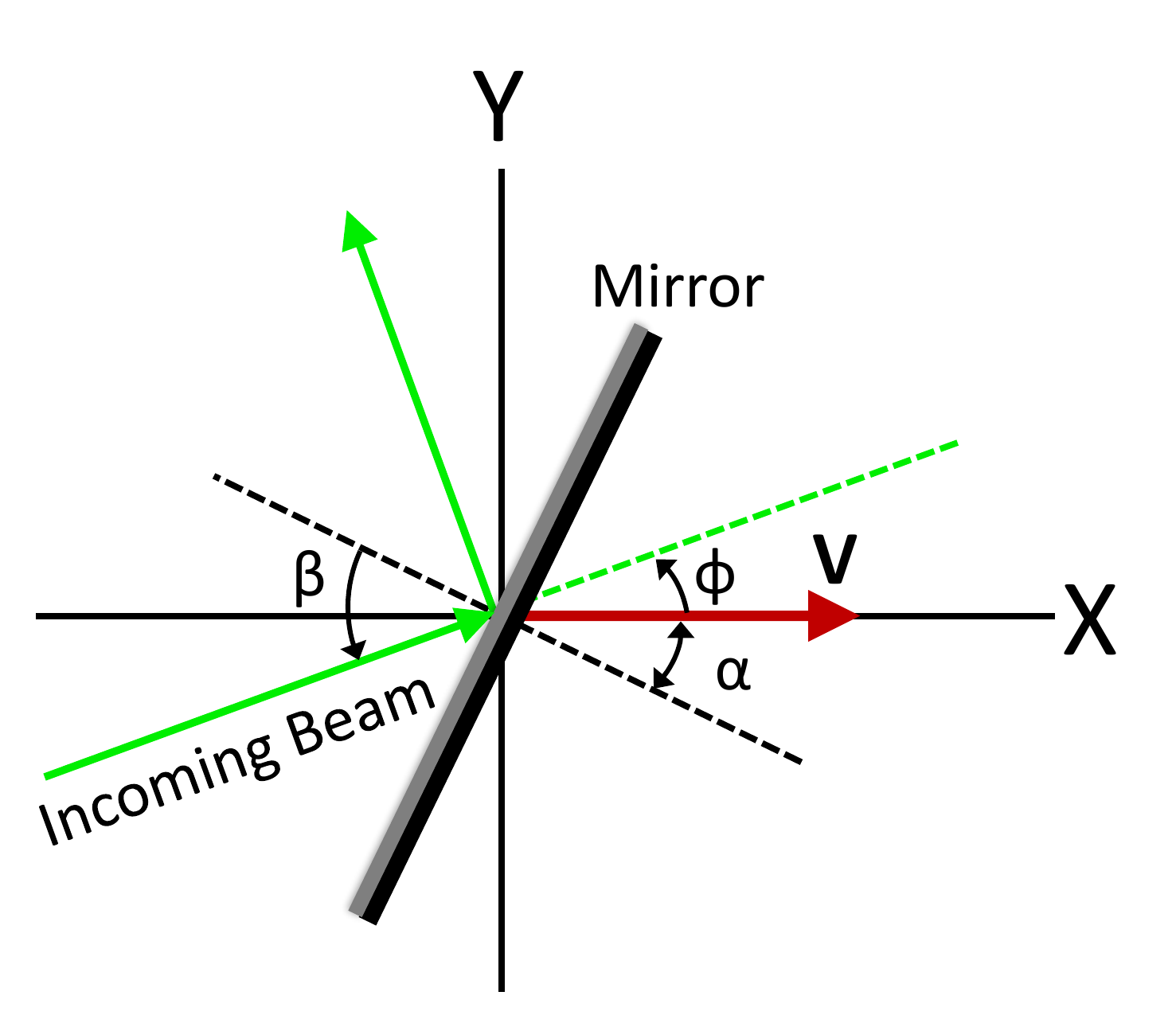}\\
\caption{ Light reflecting from a mirror moving with velocity v along the x-axis. We use the convention of Ashworth and Davies \cite{ashworth1976doppler} in which $\alpha$ is the angle between the velocity vector and the surface normal and $\phi$ is the angle between the angle of incidence (from the opposite side of the mirror) and the velocity vector.  We also define $\beta$ as the angle of incidence.  } 
\label{Mirror}
\end{figure}

\section{Theory}
We consider the Doppler shift scenario as originally proposed by Ashworth and Davies as shown in Fig. \ref{Mirror}.  A mirror is translating at constant velocity along the x-axis.  An incoming light beam with an angle of incidence $\beta$ is specularly reflected from the surface.  Using a convention by Ashworth and Davies, $\alpha$ is the angle between the surface normal and the propagation direction of the mirror and $\phi$ is the angle between the velocity vector and the angle of incidence.   

In an effort to prove that transverse Doppler shifts do not exist in reflection from a mirror, Ashworth and Davies derived a generalized special relativistic Doppler shift formula
\begin{equation}
    f_f=f_i\frac{[\tan\alpha+\frac{v}{c}\sin\phi]^2+[1-\frac{v}{c}\cos\phi]^2}{1-\frac{v^2}{c^2}+\tan^2\alpha},
\end{equation}
where $f_f$ and $f_i$ are the frequency of the light after and before the reflection from the mirror, respectively and $v$ is the speed of the mirror. In the limit $v\ll c$, we simplify the equation, namely
\begin{equation}
    \Delta f=\frac{2v}{\lambda}\frac{(\tan\alpha\sin\phi-\cos\phi)} {1+\tan^2\alpha},
\end{equation}
where $\Delta f=f_f-f_i$ and $\lambda$ is the wavelength of the light.  

Our aim is to rewrite this result only in terms of the mirror's surface normal rather than with respect to the velocity vector.  We first use the trigonometric identity $1+\tan^2\alpha=\sec^2\alpha=1/\cos^2\alpha$, which allows us to write 
\begin{equation}
    \Delta f=\frac{2v}{\lambda}(\sin\alpha\cos\alpha\sin\phi-\cos^2\alpha\cos\phi).
\end{equation}
We now make a change of variables.  Instead of using $\phi$ (the angular difference between the angle of incidence and the velocity vector), we directly use the angle of incidence $\beta$ given by $\beta=\phi+\alpha$.  This implies 
\begin{equation}
    \Delta f=\frac{2v}{\lambda}\left[\sin\alpha\cos\alpha\sin(\beta-\alpha)-\cos^2\alpha\cos(\beta-\alpha)\right].
\end{equation}
Using the angle sum relations $\cos(\beta-\alpha)=\cos\beta\cos\alpha+\sin\beta\sin\alpha$ and $\sin(\beta-\alpha)=\sin\beta\cos\alpha-\cos\beta\sin\alpha$ as well as $\sin^2\alpha+\cos^2\alpha=1$, we arrive at the relation 
\begin{equation}
    \Delta f=-\frac{2v}{\lambda}\cos\beta\cos\alpha.
    \label{MainResult}
\end{equation}

This formula is more intuitive than the original Ashworth-Davies result and easier to use experimentally in describing a physical system as all angles are defined with respect to the mirror's surface normal. The formula contains the familiar Einstein formula \cite{einstein1905elektrodynamik}, in the low velocity limit, for $\alpha=0$ (the scenario where the mirror surface normal also lies on the x-axis).  

\begin{figure}[htp]
\includegraphics[width=.45\textwidth]{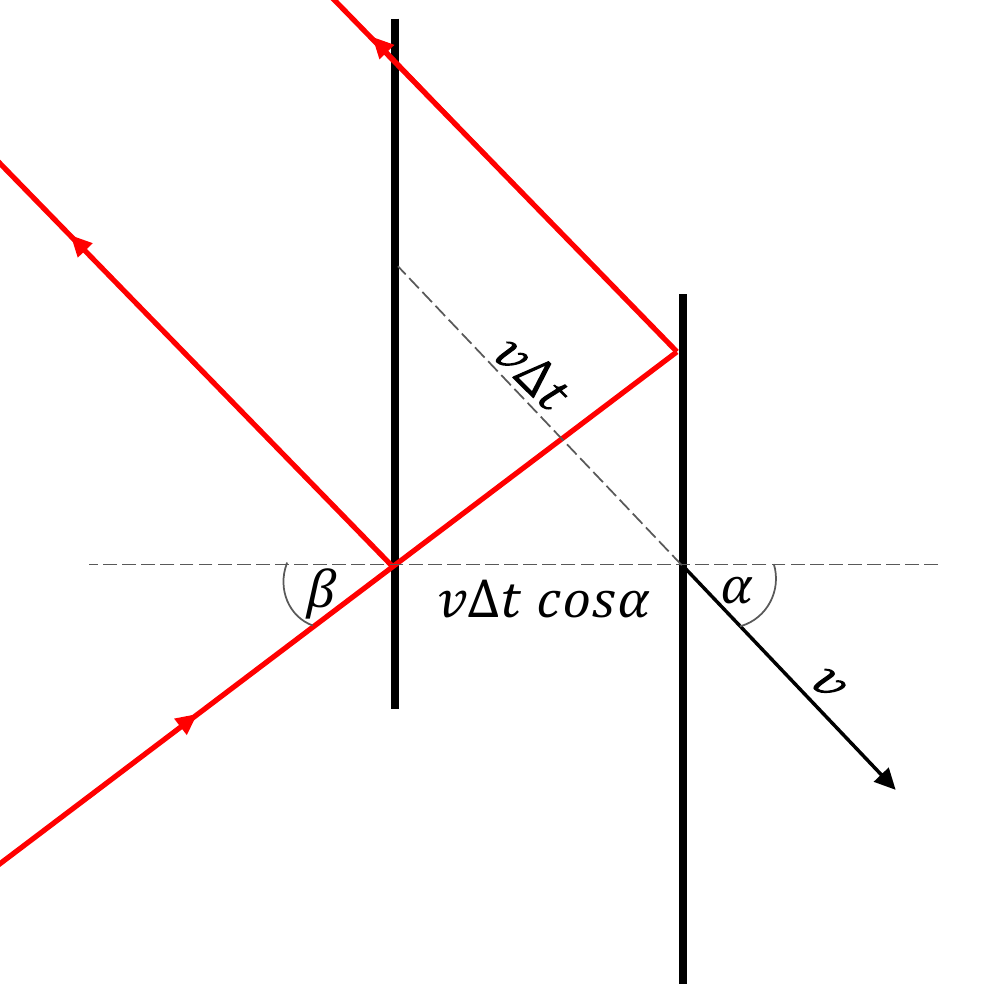}\\
\caption{Consider a mirror moving at nonrelativistic speed $v$ in an angle $\alpha$ relative to its surface normal, during a time interval $\Delta t$. The mirror velocity in the direction of its normal is $v\cos \alpha$, so the distance between the mirror before and after the time interval will be $v \cos \alpha \Delta t$. Repeating Gjurchinovski derivation in \cite{gjurchinovski2005doppler} but replacing $v \Delta t$ with $v \cos \alpha \Delta t$ will yield Eqn. \ref{MainResult}  } 
\label{wavefronts}
\end{figure}

The result is also in good agreement with a simple addition to the intuitive derivation by Gjurchinovski \cite{gjurchinovski2005doppler}.  Gjurchinovski considered sequential wavefronts reflecting with temporal separation $\Delta t$ from a mirror moving at velocity $v$ parallel to the surface's normal.  The first wavefront reflected from the mirror at the first time and after the time interval $\Delta t$ the mirror reflected the second wavefront. During the time interval $\Delta t$ the mirror propagated a distance $v\Delta t$. The distance between those two surfaces (the mirror surface at two different times) was $v\Delta t$ if the mirror propagated in the direction of the surface normal. Using only the constancy of the speed of light in a vacuum with this assumption, and without using a Lorentz transform, he reproduces Einstein's equation \cite{einstein1905elektrodynamik}: 
\begin{equation}
    f_f=f_i\frac{1 - 2 \frac{v}{c}\cos\beta + \frac{v^2}{c^2} }{1-\frac{v^2}{c^2}},
    \label{einstein}
\end{equation}

However, we note that if the velocity of the mirror were not in the direction of the surface normal, but at an angle $\alpha$, the mirror would have only moved $v\Delta t \cos\alpha$ between sequential wavefronts, as seen in Fig \ref{wavefronts}. If we only care about the velocity in order to calculate the distance between the mirror before and after a time interval $\Delta t$ and consider only nonrelativistic velocities (and thus ignoring length contraction), we can easily replace $v$ with $v \cos \alpha$ and ignore second order of $\frac{v}{c}$ in Eqn. \ref{einstein}, thus reproducing Eqn. \ref{MainResult}  

\begin{figure}[htp]
\includegraphics[width=.51 \textwidth]{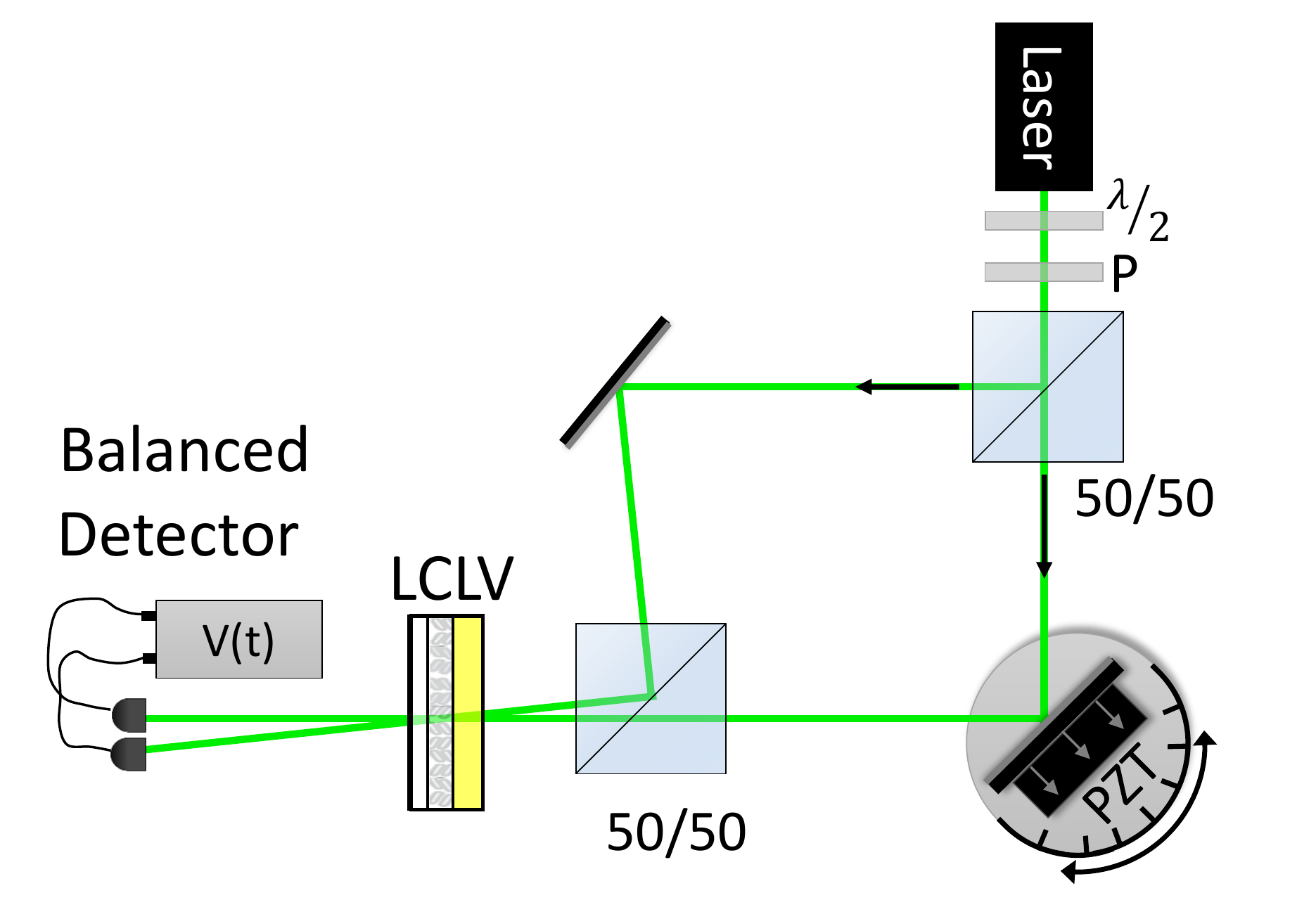}\\
\caption{ The experimental setup. A displaced Mach-Zehnder interferometer, where one of the mirrors is mounted to an adjustable moving platform, controlled by a piezo-electric crystal (PZT). A slight angle between the beams is fabricated in order to create a wave mixing where the beams meet at the Liquid Crystal Light Valve (LCLV). The two primary diffracted output orders of the LCLV are focused on a balanced detector. } 
\label{Setup}
\end{figure}

\section{Experimental Setup}
To experimentally test these results, we use a liquid crystal light valve (LCLV) \cite{bortolozzo2013precision} (see Supplementary Materials for more information).  While any experimental system that can precisely measure Doppler shifts can be used, we find the LCLV to be exceptionally ideal. As shown in \cite{bortolozzo2013precision}, a LCLV can measure down to $\mu Hz/Hz^{1/2}$ meaning that it is several orders of magnitude more sensitive than any other system per measurement time.  This means that the movement of the mirror can be made to be very small, which has the effect of minimizing alignment issues as we rotate through the various measurement angles.

The experimental setup is shown in Figure \ref{Setup}. Collimated light from a laser at 532 nm is split on a 50/50 beamsplitter. One of the mirrors is mounted to an adjustable moving platform that is controlled by a piezoactuator (PZT) so that the beam that's reflected from the mirror experiences a Doppler shift relative to the other beam. A Mach-Zehnder type setup is used to create a slight angle between the two beams on the order of 0.01 radians, and both beams are then incident on a LCLV. A small angle between the beams is needed in order to create two-wave mixing in the Raman-Nath regime of the LCLV (see Supplementary Materials for more information). The two primary diffracted output orders of the LCLV are focused on a balanced detector. The difference in the intensities of the two beams hitting the balance detector is proportional to the difference in frequency between the two beams, $\Delta f$ \cite{bortolozzo2013precision}.  

To test the Doppler shift dependence on the direction of propagation of the mirror, the PZT direction was changed while the angle of incidence was fixed.  With each iteration of the experiment, the moving platform was tuned to a different angle, and the mirror sitting on it was tuned to be with a fixed angle relative to the incoming beam at $\beta = 45 ^{\circ}$.  This caused the movement of the platform to be in a different direction relative to the mirror's surface normal (which corresponds to $\alpha$ as defined above). The PZT was driven by an arbitrary waveform generator producing triangle waves at 20 mHz and a peak-to-peak voltage of 15V.  The PZT response was measured to be approximately 60 nm/V.  The experiment was repeated 36 times, with alpha ranging from $-180^{\circ}$ to $-180^{\circ}$. The absolute value of the mirror's velocity was constant through all the iterations and equal to $V= 36$ nm / s.  

The errors in the experiment are primarily from seismic and acoustic vibrations as well as from fluctuating air currents. While we had various forms of active and passive noise reduction, the light valve is particularly sensitive to noise.  Further, the experiments are run at slow speeds meaning that 1/f noise is significant.

\begin{figure}[htp]
\includegraphics[width=.45\textwidth]{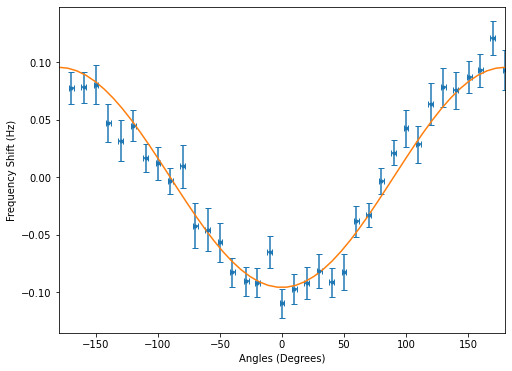}\\
\caption{ Experimental results. The Doppler signal amplitude vs velocity angle $\alpha$ relative to the surface normal is shown in the blue points along with the error bars. We used a fixed angle of incidence of $\beta=45^\circ$. The orange curve is the cosine curve from Eq. \ref{MainResult}, when plugging the parameters of our experiment ($\lambda = 532$ nm, $V = 36$ nm/s). } 
\label{CosineData}
\end{figure}

\section{Results}

The experimental results are shown in Fig. \ref{CosineData}. The blue data points are the measured amplitude of the Doppler shift from the moving mirror. The X-axis is the angle between the direction of the mirror's propagation to the mirror's normal ($\alpha$). The orange curve is a least-squares fitted cosine function.  Hence the experimental results are in excellent agreement with the main result of this paper (Eqn. \ref{MainResult}).  We have thus shown that there is not only a cosine term that arises from the angle of incidence (measured relative to the surface normal) but also from the direction of mirror propagation relative to the surface normal.    

\section{Discussion and Conclusion}
We have shown that there is a simple nonrelativistic Doppler correction to the frequency of light that accounts for both the angle of incidence and the angle of mirror propagation relative to the surface normal.  We showed that the more cumbersome Ashworth-Davies result simplifies to this result in this limit.  We further showed using precision differential frequency measurements that the cosine behavior is observed.  
These results provide a novel simplified form of the Doppler shift in the nonrelativistc limit and can have a large impact in practical domains of application.


We gratefully acknowledge support from the Hebrew University.

\vfill{\eject}

\bibliography{Sagnac}

\end{document}


\section {LCLV description}
The LCLV is a device composed of a liquid crystal (LC) layer placed between a photorefractive \(B_{12}SiO_{20}\) (BSO) crystal substrate, here used because of its large photoconductivity in the visible range \cite{bortolozzo2009PRA}, and a glass window. Transparent electrodes of {\it Indium-Tin-Oxide (ITO)} are deposited  over the BSO and the glass window, and they allow the application of an external bias voltage $V_0$ across the LC layer. 
The LC used in our LCLV is a planar aligned nematic liquid crystal layer of E48 (from Merck). The extraordinary and ordinary index are, respectively, $n_e \sim 1.75$ and $n_o \sim 1.52$ at 589nm, for a total birefringence $n_e-n_o \sim 0.23$. The thickness of the LC layer is $d=15$ $\mu m$.

When the external voltage is applied, the LC molecules tend to align in such a way to become parallel to the direction of the applied field. Because of their birefringence, the refractive index changes accordingly.  
Since the BSO adjusts the effective voltage $V_{LC}$ across the LC as a function of the incident light intensity, in the light valve, the LC orientation, and thus the refractive index depends on the local illumination in the photoconductive layer \cite{bortolozzo2009PRA}. This property is used to achieve spatial light modulation of an incoming reading beam depending on the writing illumination on the photoconductive side of the light valve. In a similar way, two-beam coupling can be realized by sending two interfering beams on the LCLV, thus, creating a fringe pattern on the photoconductor and a corresponding refractive index grating in the LC layer. During this process, also called two-wave mixing, the beams exchange phase and amplitude information thanks to their mutual diffraction over the refractive index grating. The mechanism for the two-wave mixing, and the related detection of the Doppler shift, are described in the next section.

\section {Two-wave mixing and Doppler shift detection}
When two optical beams are incident on the LCLV, they create an intensity fringe pattern:
$$A_1 e^{\imath (\Vec{k}_1 \vec{r} - 2\pi f_1 t )}+A_2 e^{\imath (\Vec{k}_2 \vec{r} - 2\pi f_2 t )}= |A_1|^2+|A_2|^2+A_1 A_2^\star e^{\imath(\vec{K}_g-2\pi f_D t)}+c.c.$$

where $A_1$ and $A_2$ are the amplitudes, \(\Vec{k}_1\) and \(\Vec{k}_2\) are the wave vectors of the two optical beams respectively. 
The frequency detuning of the two beams is defined as $f_D=f_1-f_2$, where $f_1$ and $f_2$ are the frequencies of the two beams respectively. In our case, $f_D$ represent the Doppler shift produced by the two beams travelling over the two open-path loops of the interferometer.
The grating wave vector is \( \Vec{K}_g= \Vec{k}_1-\Vec{k}_2\). 

The interference pattern induces in the LCLV a refractive index grating with a period of \(\Lambda\equiv{2\pi}/K_g\) that allows an energy exchange between the two optical beams. In our experiment the grating period is greater than the thickness of the LC, $d$, so the beam coupling occurs in the Raman-Nath regime of diffraction \cite{Yarivbook}.

This two-wave mixing process has been shown to lead to a very slow group velocity \cite{PhysRevLett.100.203603};  the slow response means that this process is nearly independent of the temporal properties of the light, which produces this effect, making the detection independent of the relative phase between the two beams.

The two output intensities are given by \cite{bortolozzo2009PRA}
$$I_{1} = |J_{0}(\rho) A_1 + iJ_{1}(\rho)e^{-i\psi} A_2|^{2}, \;\;\;I_{2} = |J_{0}(\rho) A_2 + iJ_{1}(\rho)e^{i\psi} A_1|^{2}$$

where $J_{m}$ is the Bessel function of the first kind, 
\begin{equation}
    	\rho = \frac{2k_{0}d n_{2}}{\sqrt{(1+{l}_{D}^2{K}_{g}^2)^2+(2 \pi f_{D}\tau)^2}}\,|A_1 A_2|\label{eqn:def-rho},
    \end{equation}
 is the grating amplitude and
    
    \begin{equation}
    	\tan{\psi} = \frac{ 2 \pi f_{D}\tau}{1+{l}_{D}^2{K}_{g}^2}\label{eqn:def-psi}.
    \end{equation}
Here $k_{0}$ is the optical wave-number, $n_{2}$ is the equivalent Kerr-like nonlinear coefficient, and $l_{D}$ is  the transverse diffusion length,  governed by the material constants and the applied bias voltage \cite{bortolozzo2009PRA}.
By assuming two balanced input intensities, $I_{in}=|A_1|^2=|A_2|^2$, in the case of small Doppler shift, $f_{D}\tau \ll (1+(l_{D}k_{g})^{2})$ , the measured differential intensity after the two wave mixing process is given by,
\begin{equation}
	\Delta I= I_1-I_2=I_{in}\frac{8 \pi \tau J_{0}(\rho)J_{1}(\rho)}{1+(l_{D}K_{g})^{2}} f_{D}.\label{eqn:doppler-approx}
\end{equation}
We observe that $\Delta I$ is linear with the Doppler frequency. Moreover, as noted earlier, the amplitude of our signal does not depend on the relative phase between the two beams. These features, the large linear response and the insensitivity in phase, are the main advantages of using the LCLV as the nonlinear medium to perform the two-wave mixing process.

\section{Physical parameters of the device}
The main physical parameters characterizing the LCLV and the two-wave mixing model are the transverse diffusion length $l_D$, the Kerr-like coefficient $n_2$ and the response time $\tau$. The values of these parameters are determined by the physical constants of the LC layer and the photoconductive substrate of the LCLV, as well as by the amplitude of the applied bias voltage and by the illumination conditions, wavelength and power of the beams.

The diffusion length is given by
\begin{equation} l_D=\frac{d \sqrt{\Delta \epsilon /K}}{V_{LC}}\end{equation}
where $\Delta \epsilon$ and $K$ are, respectively, the dielectric anisotropy and the elastic
constant of the liquid crystal and $V_{LC}$ is the bias voltage across the liquid crystal layer.
For the parameters used in the experiment we have $l_D\approx 20$ $\micro m$.

The Kerr coefficient $n_2$ is determined by measuring the two wave mixing gain for small intensity of the interacting beams ($I_1,I_2 <10$ $mW/cm^2$). This is done simply by switching on and off one of the input beams and by measuring the second one with a photo-detector. Indeed, the ratio between the intensities of the output beam $I_1$ over $I_{1{|I_2=0}}$ when the second beam $I_2$ is shut off is given by
\begin{equation} \frac{I_1}{I_{1{|I_2=0}}}=\frac{|J_0(\rho)A_1+\imath J_1(\rho)A_2|^2}{|A_1|^2},\end{equation}
and can, therefore, be related to $n_2$ via the $\rho$ dependency, Eq.(\ref{eqn:def-rho}).
By inserting the values used in our experiment, $|A_1|^2=4.62$ $mW/cm^2$ and $|A_2|^2=4.66$ $mW/cm^2$, we obtain $n_2\approx 0.22$ $cm^2/W$.

Then, by measuring the transient decay time of $I_1(t)$ when $I_2$ is turned off we evaluate the response time $\tau\approx 78$ $ms$.
The table here below summarizes the measured parameters values. 

\begin{center}
\begin{tabular}{ | m{2.5cm} | m{2.5cm}| m{2.5cm} | } 
  \hline
  $l_D$& $n_2$  & $\tau$ \\ 
  \hline
  $20$ $\mu m$ & $0.22$ $cm^2/W$  & $78$ $ms$ \\ 
  \hline
 \end{tabular}
\end{center}

\section {Shot noise detection limit}
The minimum detectable Doppler shift imposed by the shot noise is given by \cite{bortolozzo2013precision}:
\begin{equation}f_{SD} = \frac{1}{8 \pi \tau J_0(\rho)J_1(\rho)}\sqrt{\frac{h c}{\lambda \eta T P_{PD}}},
\label{min_Doppler}, \end{equation}
where $\eta$ and $P_{PD}$ are, respectively, the quantum efficiency and the total optical power detected by the balanced photodiodes.  
$T$ is the integration time of the detection.
The minimum detectable rotation speed can be calculated as :
\begin{equation}\Omega_{SD} =\frac{\lambda}{L \epsilon} f_{SD}= \frac{1}{8 \pi \tau J_0(\rho)J_1(\rho)L \epsilon}\sqrt{\frac{h c \lambda}{\eta T P_{PD}}},\label{min-omega} \end{equation}
where $\lambda=532$ $nm$ is the optical wavelength, $L^2$ is the area enclosed by the interferometer and $\epsilon$ is the angular difference between the beams exiting the interferometer.

We insert in Eq.(\ref{min-omega}) some values used in the experiment, $L=0.1$ $m$ and $\epsilon=7.2$ $mrad$, and we calculate $f_{SD}$ by using Eq.(\ref{min_Doppler}) with the following experimental parameters: input beam intensities $|A_1|^2=4.62$ $mW/cm^2$ and $|A_2|^2=4.66$ $mW/cm^2$, LCLV diffusion length $l_D=20$ $\mu m$, nonlinear coefficient $n_2=0.22\cdot10^{-4}$ $m^2/W$ and response time $\tau=78$ $ms$, $d=15$ $\mu m$ thickness of the LC layer. 

The measured output powers at the exit of the LCLV are a few mW, we take for example $P_{PD}=5$ $mW$, and by taking a quantum efficiency for the photodetector $\eta=1$, we find a minimal detectable Doppler shift $f_{SD}\simeq 50\,\,nHz\,{Hz}^{-1/2}$.  In the paper, we used $f_{SD}\simeq 100\,\,nHz\,{Hz}^{-1/2}$ in the paper as an estimate.

\bibliography{Sagnac}